\documentclass[preprint,authoryear,12p]{elsarticle}

\usepackage{epsfig}
\usepackage{amssymb}

\journal{Advances in Space Research}
\begin{document}
\begin{frontmatter}
\title{Spin-isospin transitions in chromium isotopes within the quasiparticle random phase approximation}
\tnotetext[footnote1]{This template can be used for all publications
in Advances in Space Research.}

\author{Sadiye CAKMAK}
\address{Department of Physics, Celal Bayar University,
Manisa, Turkey} \ead{sadiyecakmak@hotmail.com}

\author{Jameel-Un NABI}
\address{Faculty of Engineering Sciences,
GIK Institute of Engineering Sciences and Technology, Topi 23640,
Swabi, Khyber Pakhtunkhwa, Pakistan} \ead{jameel@giki.edu.pk}

\author{Tahsin BABACAN}
\address{Department of Physics, Celal Bayar University,
Manisa, Turkey}
 \ead{tahsin.babacan@cbu.edu.tr}

\author{Ismail MARAS}
\address{Department of Physics, Celal Bayar University,
Manisa, Turkey}
 \ead{ismail.maras@cbu.edu.tr}

\begin{abstract}
Beta decay and electron capture on isotopes of chromium are
advocated to play a key role in the stellar evolution process. In
this paper we endeavor to study charge-changing transitions for 24
isotopes of chromium ($^{42-65}$Cr). These include neutron-rich and
neutron-deficient isotopes of chromium. Three different models from
the QRPA genre, namely the pn-QRPA, the Pyatov method (PM) and the
Schematic model (SM), were selected to calculate and study the
Gamow-Teller (GT) transitions in chromium isotopes. The SM was
employed separately in the particle-particle ($pp$) and $pp$ +
particle-hole ($ph$) channels. To study the effect of deformation,
the SM was first used assuming the nuclei to be spherical and later
to be deformed. The PM was used both in $pp$ and $pp$ + $ph$
channels but only for the case of spherical nuclei. The pn-QRPA
calculation was done by considering both $pp$ and $ph$ forces and
taking deformation of nucleus into consideration. A recent study
proved this version of pn-QRPA to be the best for calculation of GT
strength distributions amongst the QRPA models. The pn-QRPA model
calculated GT distributions had low-lying centroids and small widths
as compared to other QRPA models. Our calculation results were also
compared with other theoretical models and measurements wherever
available. Our results are in decent agreement with previous
measurements and shell model calculations.
\end{abstract}


\end{frontmatter}

\parindent=0.5 cm

\section{Introduction}
Gamow- Teller (GT) transitions play an important role in many
astrophysical events in universe. As soon as the iron core of a
massive star exceeds the Chandrasekhar mass limit, the degenerated
pressure of electron gas can no longer support the core, and the
core starts to collapse. During early stages of collapse many
important nuclear processes, such as $\beta$ decays, electron
captures, neutrino absorption and inelastic scattering on nuclei,
appear. These reactions are mainly governed by GT (and Fermi)
transitions. GT transitions for $fp$-shell nuclei are considered
very important for supernova physics \citep{Ful85}. The GT
transitions, in $fp$-shell nuclei, play decisive roles in
presupernova phases of massive stars and also during the core
collapse stages of supernovae (specially in neutrino induced
processes). At stellar densities $\sim$ 10$^{11}$ gcm$^{-3}$, for
$fp$-shell nuclei, the electron chemical potential approaches the
same order of magnitude as the nuclear Q-value. Under such
conditions, the $\beta$-decay rates are sensitive to the detailed GT
distributions. Consequently centroids and widths of the calculated
GT distributions function become important. For still higher stellar
densities, the electron chemical potential is much larger than
nuclear Q-values. Electron capture rates become more sensitive to
the total GT strength for such high densities. To achieve a better
understanding of these notoriously complex astrophysical phenomena,
a microscopic calculation of  GT strength distributions (along with
the total strength, centroid and width of these distributions) is in
order.

The GT transitions are extremely sensitive to two-body interactions.
The GT excitations deal with the spin-isospin degree of freedom and
are executed by the $\sigma\tau_{\pm,0}$ operator, where $\sigma$ is
the spin operator and $\tau_{\pm,0}$ is the isospin operator in
spherical coordinates. The plus sign refers to the GT$_+$
transitions where a proton is changed into a neutron (commonly
referred to as electron capture or positron decay). On the other
hand, the minus sign refers to GT$_-$ transitions in which a neutron
is transformed into a proton ($\beta$-decay or positron capture).
The third component GT$_{0}$ is of relevance to inelastic
neutrino-nucleus scattering for low neutrino energies and would not
be considered further in this manuscript. Total GT$_-$ and GT$_+$
strengths (referred to as $B(GT)_-$ and $B(GT)_+$, respectively, in
this manuscript) are related by Ikeda Sum Rule as
$B(GT)_{-}-B(GT)_{+}=3(N-Z)$, where $N$ and $Z$ are numbers of
neutrons and protons, respectively \citep{Ike63}. Given nucleons are
treated as point particles and two-body currents are not considered,
the model independent Ikeda Sum Rule should be satisfied by all
calculations.

Not many measurements of GT strength in chromium isotopes have been
performed to the best of our knowledge. The decay of $^{46}$Cr was
first studied by \citep{Zio72}, who used the $^{32}$S($^{16}$O,2n)
reaction to produce $^{46}$Cr. Onishi and collaborators \citep{On05}
observed the $\beta$-decay of $^{46}$Cr to the 1$^{+}_{1}$ state  at
993 keV excitation energy in $^{46}$V. The $T = 1$ nuclei decay to
the $T = 0$ and 1$^{+}$ states of daughter nuclei were called
favored-allowed GT transitions and possessed a signature small $ft$
value. The experiment was performed at RIKEN accelerator research
facility. Two sets of independent measurement of $B(GT)_-$ strength
for $^{50}$Cr were also performed. Fujita et al. did a
$^{50}$Cr($^{3}$He, $t$)$^{50}$Mn measurement up to 5 MeV in
daughter \citep{Fuj11}. On the other hand Adachi and collaborators
\citep{Ada07} were able to perform a high resolution
$^{50}$Cr($^{3}$He, $t$)$^{50}$Mn measurement at an incident energy
of 140 MeV/nucleon and at 0$^{0}$ for the precise study of GT
transitions. The experiment was performed at RCNP, Japan. Owing to
high resolution the authors were able to measure $B(GT)_-$ strength
up to 12 MeV in $^{50}$Mn. At higher excitations above the proton
separation energy, a continuous spectrum caused by the quasifree
scattering appeared in the experiment.  Nonetheless there is a need
to perform more experiments to measure GT transitions in $fp$-shell
nuclei. Next-generation radioactive ion-beam facilities (e.g. FAIR
(Germany), FRIB (USA) and FRIB (Japan)) are expected to provide us
measured GT strength distribution of many more nuclei. It is also
expected to observe GT states in exotic nuclei near the neutron and
proton drip lines. However simulation of astrophysical events (e.g.
core-collapse supernovae) requires GT strength distributions ideally
for hundreds of nuclei. As such experiments alone cannot suffice and
one has to rely on reliable theoretical estimates for GT strength
distributions.

Isotopes of chromium are advocated to play an important role in the
presupernova evolution of massive stars. Aufderheide and
collaborators \citep{Auf94} searched for key weak interaction nuclei
in presupernova evolution. Phases of evolution in massive stars,
after core silicon burning, were considered and a search was
performed for the most important electron capture and $\beta$-decay
nuclei for 0.40 $\le Y_{e} \le$ 0.50 ($Y_{e}$ is lepton-to-baryon
fraction of the stellar matter). The rate of change of $Y_{e}$
during presupernova evolution is one of the keys to generate a
successful explosion. From their calculation, electron captures on
$^{51-58}$Cr and $\beta$-decay of $^{53,54,55,56,57,59,60}$Cr were
found to be of significant astrophysical importance, regarding their
abundance and weak rates, to control $Y_{e}$ in stellar matter.
Heger and collaborators \citep{Heg01} performed simulation studies
of presupernova evolution employing shell model calculations of
weak-interaction rates in the mass range A = 45 to 65. Electron
capture  rates on $^{50,51,53}$Cr were found to be crucial for
decreasing the $Y_{e}$ of the stellar matter. Similarly, it was
shown in the same study that $\beta$-decay rates of
$^{53,54,55,56}$Cr played a significant role in increasing the
$Y_{e}$ content of the stellar matter. There exists a strong
connection between supernova explosion and the $r$-process
nucleosynthesis \citep{Cow04}. These and similar studies of
presupernova evolution provided us the motivation to perform a
detailed study of GT transitions for chromium isotopes. In this work
we calculate and study GT distributions of twenty-four (24) isotopes
of chromium, $^{42-65}$Cr, both in the electron capture and
$\beta$-decay direction.

The theoretical formalism used to calculate the GT strength
distributions in the PM, SM and pn-QRPA models is described briefly
in the next section. We compare our results with other model
calculations and measurements in Section 3. A decent comparison
would put more weight in the predictive power of the QRPA models
used in this work. It is pertinent to mention again that our
calculation includes many neutron-rich and neutron-deficient
isotopes of chromium for which no experimental data is available for
now. Core-collapse simulators rely heavily on reliable theoretical
estimates of the corresponding weak rates in their codes. Section 4
finally summarizes our work and states the main findings of this
study.

\section{Theoretical Formalism}
GT strength distributions of $^{42-65}$Cr isotopes were calculated
 by using three different (microscopic) quasiparticle
random phase approximation (QRPA) models namely the pn-QRPA
(\citep{Sta90, Hir93}), the Pyatov Method (PM) and the Schematic
Model (SM) (for details see \citep{Pya77,Bab04,Bab05,Bab05a}). The
QRPA treats the particle-particle ($pp$) and hole-hole amplitudes in
a similar way as in particle-hole ($ph$) amplitudes.  The QRPA takes
into account pairing correlations albeit in a non-perturbative way.
Earlier a similar study  for calculation of GT transitions for key
titanium isotopes, using the same models, was performed
\citep{Cak14}. It was shown in Ref. \citep{Cak14} that the
pn-QRPA~(C) (incorporating $pp$ and $ph$ forces as well as
deformations) was a far better model than pn-QRPA~(A) (basic model
in which only the interaction in ph channel was considered) and
pn-QRPA~(B) (including both $pp$ and $ph$ forces but no
deformation). Further the pn-QRPA (C) was the only model that was
satisfying the Ikeda Sum Rule. On the other hand, the SM and PM
showed satisfying results in their respective subcategories for
calculation of GT transitions \citep{Cak14}. Based on the results of
\citep{Cak14}, we decided to categorize the PM and SM models in two
categories as Model~(A) and Model~(B). Model~(A) is spherical model
with \textit{only} $ph$ channel, while Model~(B) is also spherical
with \textit{both} $ph$ and $pp$ interactions. The $pp$ interaction
is usually thought to play a less important role in $\beta^-$ decay
but shown to be of critical importance in $\beta^+$ decay and
electron capture reactions \citep{Hir93}. GT calculations have
demonstrated deformation parameter to be one of the most important
parameters in pn-QRPA calculations \citep{Ste04}. The SM was also
employed with $pp$ + $ph$ interactions \textit{and} deformation of
nucleus and is referred to as SM~(C) model in this work. The pn-QRPA
model was not categorized (for reasons mentioned above) and
incorporated $pp$ + $ph$ interactions as well as nuclear
deformation. In the PM, efforts for incorporation of deformation is
currently in progress and it is hoped that results would be reported
in near future. To summarize GT calculation for twenty-four chromium
isotopes, both in electron capture and $\beta$-decay direction, was
performed using six different models: SM~(A), SM~(B), SM~(C),
PM~(A), PM~(B) and pn-QRPA in this work.

In this section, we give necessary formalism used in the PM, SM and
pn-QRPA models. Detailed formalism may be seen in \citep{Cak14} and
is not reproduced here for space consideration.

\subsection{Pyatov Method (PM) and the Schematic Model (SM)}
The SM Hamiltonian for GT excitations in the quasi particle
representation is given as
\begin{eqnarray}
H_{SM}=H_{SQP}+h_{ph}+h_{pp}, \label{Eqt. 1}
\end{eqnarray}
where $H_{SQP}$ is the Single Quasi Particle (SQP) Hamiltonian,
$h_{ph}$ and $h_{pp}$ are the GT effective interactions in the $ph$
and $pp$ channels, respectively \citep{Nec10}. The effective
interaction constants in the $ph$ and $pp$ channel were fixed from
the experimental value of the Gamow-Teller resonance (GTR) energy
and the $\beta^{±}$-decay \textit{log(ft)} values between the low
energy states of the parent and daughter nucleus, respectively. In
order to restore the super symmetry property of the pairing part in
total Hamiltonian, certain terms, which do not commute with the GT
operator, were excluded from the total Hamiltonian and the broken
commutativity of the remaining part due to the shell model mean
field approximation was restored by adding an effective interaction
term $h_{0}$:
\begin{eqnarray}
[H_{SM}-h_{ph}-h_{pp}-V_{1}-V_{c}-V_{ls}+h_{0},G_{1\mu}^{\pm}]=0,
\label{Eqt. 2}
\end{eqnarray}
or
\begin{eqnarray}
[H_{SQP}-V_{1}-V_{c}-V_{ls}+h_{0},G_{1\mu}^{\pm}]=0, \label{Eqt. 3}
\end{eqnarray}
where $V_{1}$, $V_{c}$ and $V_{ls}$ are isovectors, Coulomb and spin
orbital term of the shell model potential, respectively. The
Gamow-Teller  operators $G_{1\mu}^{\pm}$ were defined in the
following way:
\begin{eqnarray}
G_{1\mu}^{\pm}=\frac{1}{2}\sum_{k=1}^{A}[\sigma_{1\mu}(k)t_+(k)+\rho(-1)^\mu\sigma_{1-\mu}(k)t_-(k)]
      (\rho=\pm1),
\end{eqnarray}
where $\sigma_{1\mu}(k)=2s_{1\mu}(k)$ is spherical component of
Pauli operators, $t_\pm=t_x(k)\pm it_y(k)$ are the raising and
lowering isospin operators. The restoration term $h_{0}$ in
Eq.~(\ref{Eqt. 3}) was included in a separable form:
\begin{eqnarray}
h_{0}=\sum_{\rho=\pm}\frac{1}{2\gamma_{\rho}}\sum_{\mu=0,\pm1}[H_{SQP}-V_{c}-V_{ls}-V_{1},G_{1\mu}^{\rho}]^{\dagger}\cdot\nonumber
\end{eqnarray}
\begin{eqnarray}
[H_{SQP}-V_{c}-V_{ls}-V_{1},G_{1\mu}^{\rho}] \label{Eqt. 4}.
\end{eqnarray}
The strength parameter $\gamma_{\rho}$ of $h_{0}$ effective
interaction was found from the commutation condition in
Eq.~(\ref{Eqt. 3}) and the following expression was obtained for
this constant (for details see Ref. \citep{Sal06}):
\begin{eqnarray}
\gamma_{\rho}=\frac{\rho}{2}\langle0|[[H_{SQP}-V_{c}-V_{ls}-V_{1},G_{1\mu}^{\rho}],G_{1\mu}^{\rho}]
|0\rangle.\nonumber
\end{eqnarray}
The total Hamiltonian of the system according to PM is
\begin{eqnarray}
H_{PM}=H_{SQP}+h_{0}+h_{ph}+h_{pp} \label{Eqt. 5}.
\end{eqnarray}
The GT transition strengths were calculated using the formula:
\begin{eqnarray}
B_{GT}^{(\pm)}(\omega_{i})=\sum_{\mu}\langle
1_{i,\mu}^{+}|G_{1\mu}^{\pm}|0^{+}\rangle^{2},
\end{eqnarray}
where $\omega_{i}$ are the excitation energy in the nucleus. The
$\beta^{\pm}$ transition strengths were defined as
\begin{eqnarray}
B(GT)_{\pm}=\sum_{i}B_{GT}^{(\pm)}(\omega_{i}), \label{Eqt. 7}
\end{eqnarray}
and are required to fulfill the Ikeda Sum Rule (ISR)
\begin{eqnarray}
ISR=B(GT)_{-}-B(GT)_{+}\cong3(N-Z). \label{Eqt. 8}
\end{eqnarray}
Detailed mathematical formalism maybe seen in \citep{Cak14}. As
shown above, the main difference between the PM and SM models is
that the effective interaction term ( $h_{0}$) is not added to the
total Hamiltonian in the SM. The $h_{0}$ term can make width and
centroid values in the PM lower than corresponding values in SM (for
further details, see Refs. \citep{Bab05,Bab05a,Nec10,sel04,Sal03}).

\subsection{The pn-QRPA Method}
The Hamiltonian of the pn-QRPA model is given by
\begin{equation}
H^{QRPA} = H^{sp} + V^{pair} + V ^{ph}_{GT} + V^{pp}_{GT}.
\label{Eqt. 9}
\end{equation}
Single particle energies and wave functions were calculated in the
Nilsson model which takes into account nuclear deformation. Pairing
in nuclei was treated in the BCS approximation. In the pn-QRPA
model, proton-neutron residual interaction occurs through both $pp$
and $ph$ channels. Both the interaction terms were given a separable
form. For further details see Ref. \citep{Hir93}. The reduced
transition probabilities for GT transitions from the QRPA ground
state to one-phonon states in the daughter nucleus were obtained as
\begin{equation}
B_{GT}^{\pm}(\omega) = |\langle \omega, \mu
\|t_{\pm}\sigma_{\mu}\|QRPA\rangle|^{2}, \label{Eqt. 14}
\end{equation}
where the symbols have their usual meaning. $\omega$ represents
daughter excitation energies. $\mu$ can only take three values (-1,
0, 1) and represents the third component of the angular momentum.
The $\beta^{\pm}$ transition strengths were calculated as in
Eq.~(\ref{Eqt. 7}) and satisfied the ISR (Eq.~(\ref{Eqt. 8})).

For odd-A nuclei, there exist two different types of transitions:
(a) phonon transitions with the odd particle acting only as a
spectator and (b) transitions of the odd particle itself. For case
(b) phonon correlations were introduced to one-quasiparticle states
in first-order perturbation. For further details, we refer to
\citep{Hir93}.

Experimentally adopted values of the deformation parameters, for
even-even isotopes of chromium ($^{48,50,52,54}$Cr), extracted by
relating the measured energy of the first $2^{+}$ excited state with
the quadrupole deformation, were taken from \citep{Ram87}. For other
cases the deformation of the nucleus was calculated as
\begin{equation}
\delta = \frac{125(Q_{2})}{1.44 (Z) (A)^{2/3}},
\end{equation}
where $Z$ and $A$ are the atomic and mass numbers, respectively, and
$Q_{2}$ is the electric quadrupole moment taken from M\"{o}ller and
collaborators \citep{Moe81}. Q-values were taken from the recent
mass compilation of Audi and collaborators \citep{Aud03}.

\section{GT$_{\pm}$ Strength Distributions}
Our ultimate goal is to calculate reliable and microscopic weak
rates for astrophysical environments, many of which cannot be
measured experimentally. The theoretical calculation poses a big
challenge. For example, it was concluded that $\beta$-decay and
capture rates are exponentially sensitive to the location of
GT$_{+}$ resonance while the total GT strength affect the stellar
rates in a more or less linear fashion \citep{Auf96}. Weak rates,
with an excited parent state, are required in sufficiently hot
astrophysical environments. But rates from excited states are
difficult to get: an $(n,p)$ experiment on a nucleus $(Z,A)$ shows
where in $(Z-1,A)$ the GT$_{+}$ centroid corresponding only to the
ground state of $(Z,A)$ resides. The calculations, described in this
paper, are also limited only to ground state parents, although we
hope to tackle excited parents in the future. For a discussion on
calculation of excited state GT strength functions using the pn-QRPA
model we refer to \citep{Nab99, Nab99a, Nab04}.

For this paper, we focus on the variation in calculations of GT
strength distribution using different QRPA models, to give us an
idea of the theoretical uncertainty.  We multiply results of all
QRPA calculations by a quenching factor of $f_{q}^{2}$ = (0.6)$^{2}$
\citep{Vet89, Cak14} in order to compare them with experimental data
and prior calculations, and to later use them in astrophysical
reaction rates. The re-normalized Ikeda sum rule in all our models
subsequently translates to
\begin{equation}
Re-ISR = B(GT)_{-}-B(GT)_{+}\cong 3f_{q}^{2}(N-Z). \label{Eqt. ISR}
\end{equation}

We first present the calculated GT distributions using all our six
QRPA models for all twenty-four isotopes of chromium. We divide the
isotopes of chromium into even-even and odd-A categories. To save
space we only show the calculated total $B(GT)$ strength, centroid
and width of the GT strength distributions, in both electron capture
and $\beta$-decay directions, for all chromium isotopes.

\begin{figure}
\begin{center}
\includegraphics*[width=15cm,angle=0]{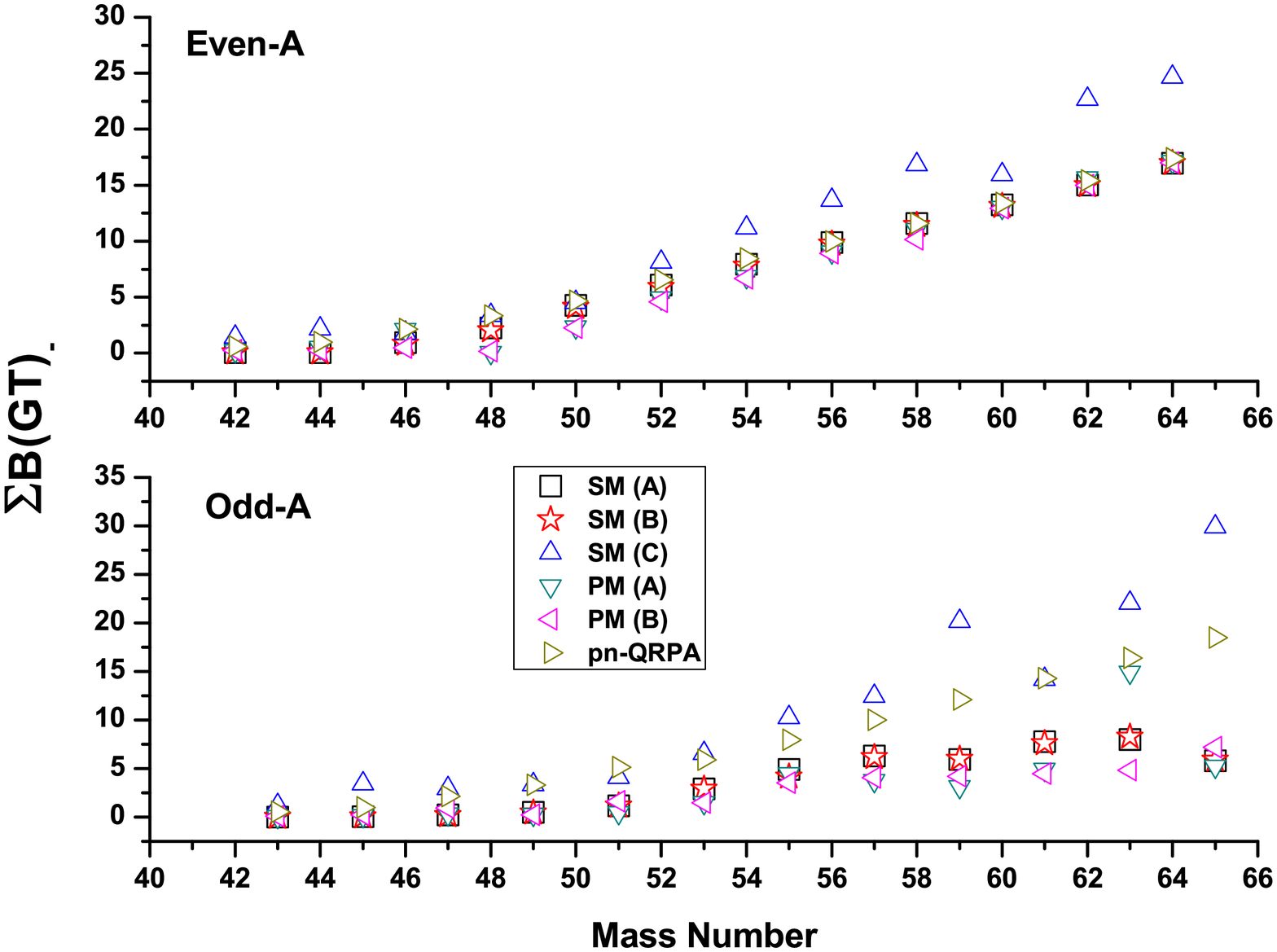}
\end{center}
\caption{Comparison of total calculated B(GT)$_{-}$ in all six QRPA
models. The upper panel shows result for even mass chromium isotopes
while the lower panel depicts result for odd mass.} \label{fig1}
\end{figure}

The total GT strength plays a very important role for the weak rates
in stellar environment. They affect the stellar rates in a more or
less linear fashion \citep{Auf96}. Fig.~\ref{fig1} compares the
calculated total $B(GT)_-$ strength values using all six QRPA
models. The upper panel shows result for even-even isotopes whereas
the lower panel those for odd-A isotopes of chromium. The total
$B(GT)_{-}$ strength increases as the mass number increases. In case
of even-~A Cr isotopes, all six models calculate similar results for
total GT strength in $\beta^-$ direction up to $^{52}$Cr isotope.
Beyond this isotope, the SM~(C) calculates bigger strength than
other models. In odd-A Cr isotopes, the SM~(C) model again shows a
tendency to calculate bigger strength values. The SM~(A), SM~(B),
PM~(A) and PM~(B) models calculate, in general, lower strength
values. Beyond $^{53}$Cr isotope, the difference between the
calculated values of the SM~(C) and the pn-QRPA starts to increase
with increasing mass number.

\begin{figure}
\begin{center}
\includegraphics*[width=15cm,angle=0]{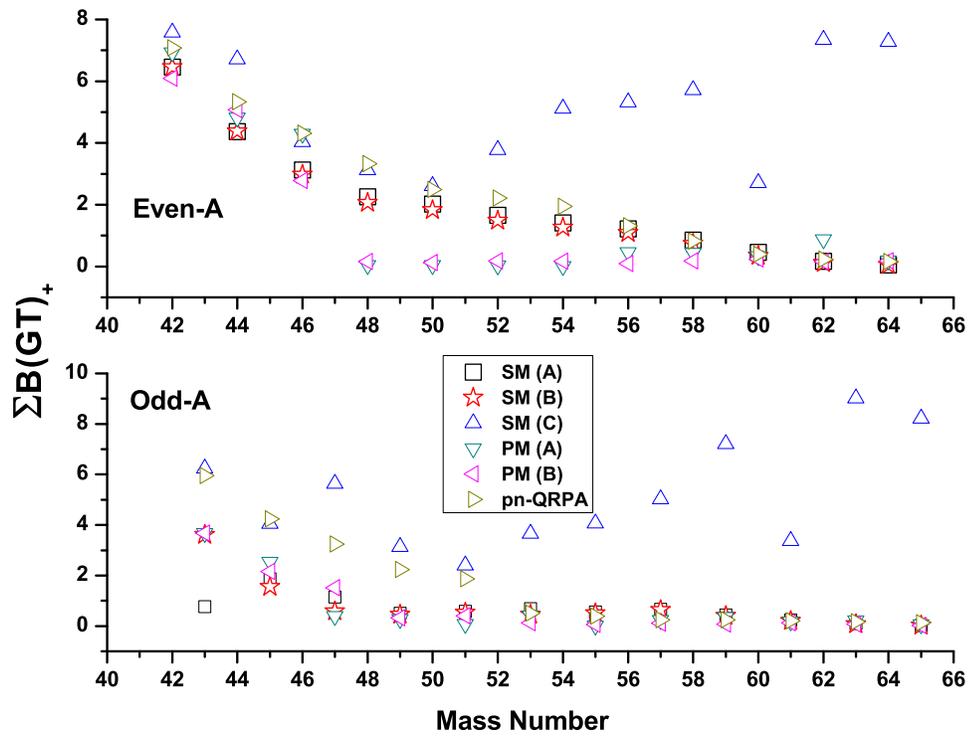}
\end{center}
\caption{Same as Fig.~\ref{fig1} but for total  B(GT)$_{+}$. }
\label{fig2}
\end{figure}

On the other hand, the total $B(GT)_{+}$ strength should decrease
monotonically with increasing mass number (separately for even-even
and odd-A cases). For even-even isotopes of chromium, this trend is
followed by the pn-QRPA, SM~(A)  and SM~(B) models
(Fig.~\ref{fig2}). For the odd-A nuclei, only the pn-QRPA model
follows this trend. It is noted that SM~(C) model has a tendency to
calculate biggest strength values whereas PM~(B) the smallest. The
difference between the two model calculations tends to increase with
mass number. For odd-A isotopes of chromium, all QRPA models, except
SM~(C), calculate almost same strength beyond $^{51}$Cr.

\begin{figure}
\begin{center}
\includegraphics*[width=15cm,angle=0]{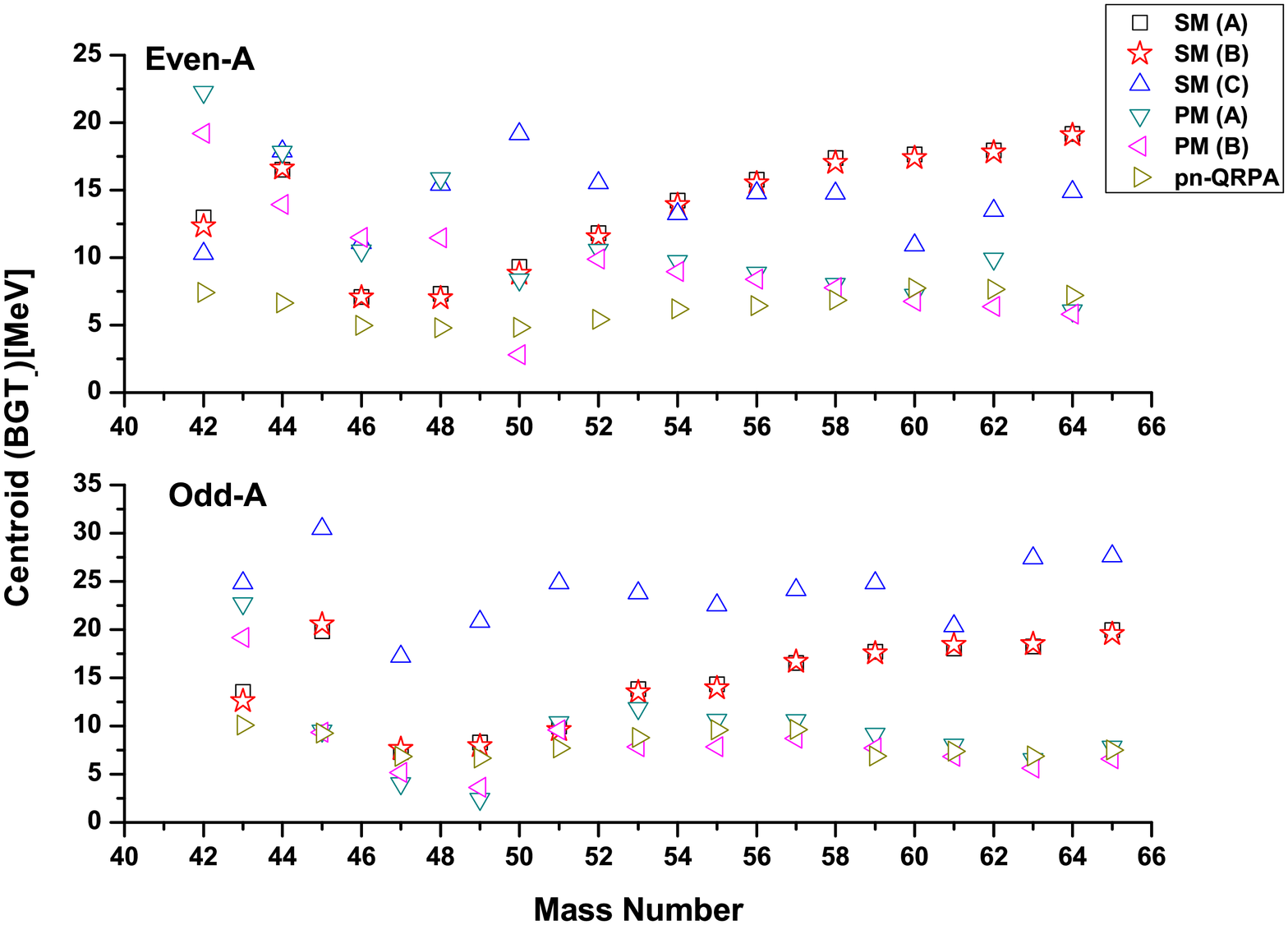}
\end{center}
\caption{Same as Fig.~\ref{fig1} but for centroid of calculated GT
distributions in $\beta$-decay direction.} \label{fig3}
\end{figure}

We next present the comparison of the calculated centroids of the GT
distributions within the six QRPA models. The $\beta$-decay and
capture rates are exponentially sensitive to the location of
GT$_{+}$ resonance \citep{Auf96} which in turn translates to the
placement of GT centroid in daughter. Fig.~\ref{fig3} shows the
calculated GT centroids, in units of MeV,  for even-even and odd-A
isotopes of chromium in the $\beta$-decay direction. The figure
shows that, in general, the SM~(C) model places the centroid at high
excitation energies in manganese isotopes whereas the pn-QRPA at
much lower energies. The other model calculations fall in between.
The inclusion of particle-particle interaction affects the centroid
results in the PM model. In general, it places the centroid at
higher excitation energies in daughter. The SM~(A) and SM~(B) models
calculate almost identical centroids. There is not much spread in
the calculated centroid values by the pn-QRPA model. The calculated
centroid values, in the case of pn-QRPA model, lies within the range
4--10 MeV for all 24 isotopes of chromium.

\begin{figure}
\begin{center}
\includegraphics*[width=15cm,angle=0]{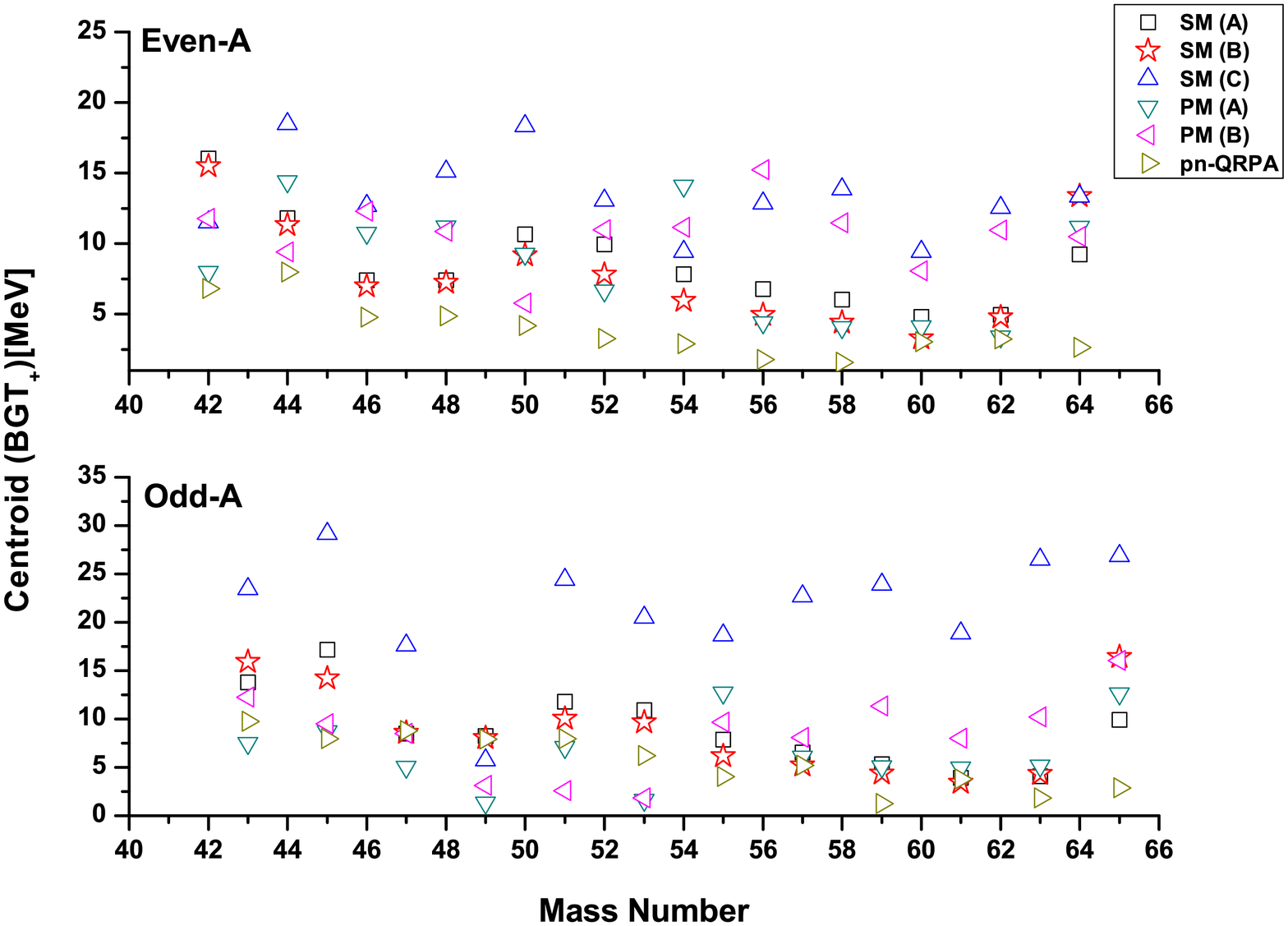}
\end{center}
\caption{Same as Fig.~\ref{fig1} but for centroid of calculated GT
distributions in electron capture direction.} \label{fig4}
\end{figure}

Fig.~\ref{fig4} displays the calculated centroids of the GT strength
distributions in the electron capture direction for all 24 isotopes
of chromium. Once again the figure shows that the SM~(C) model
possess a tendency of placing the centroids at high excitation
energy in daughter. Except for $^{42}$Cr, the deformation increases
the centroid values by more than a factor of 2--3 in the SM. The
pn-QRPA model, on the other hand, calculates low-lying centroids.
The pn-QRPA calculated centroid values are up to factor of 2--3
smaller than other models in all even Cr isotopes except for
$^{60,62}$Cr. The spread in the calculated centroid values is again
low for the pn-QRPA model, unlike other QRPA models, and lies
roughly within the range of 2--9 MeV. The lower panel of
Fig.~\ref{fig4} shows that inclusion of deformation in the SM
results in placement of centroid at higher excitation energies akin
to the case of even-even chromium isotopes.

\begin{figure}
\begin{center}
\includegraphics*[width=15cm,angle=0]{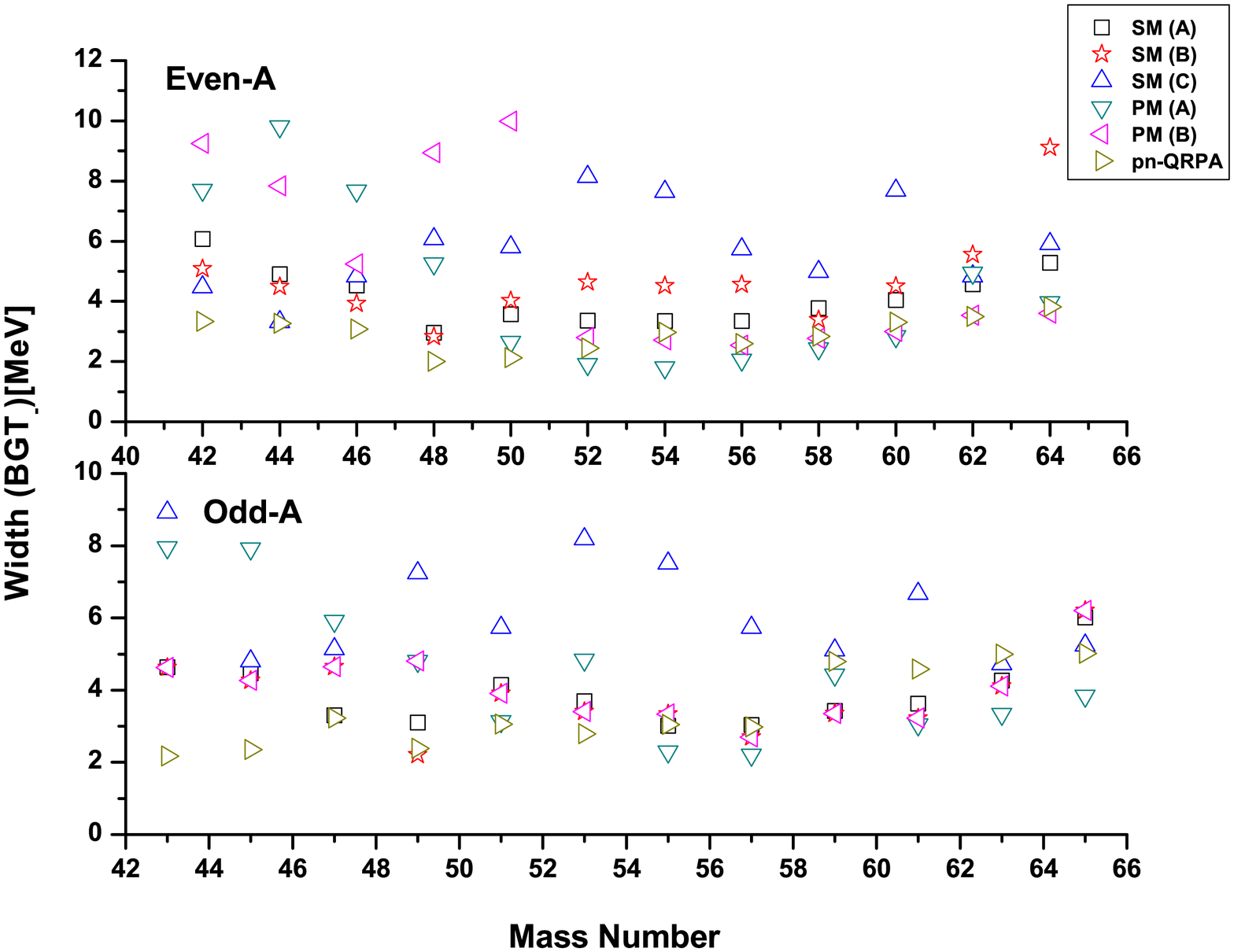}
\end{center}
\caption{Same as Fig.~\ref{fig1} but for width of calculated GT
distributions in $\beta$-decay direction.} \label{fig5}
\end{figure}

\begin{figure}
\begin{center}
\includegraphics*[width=15cm,angle=0]{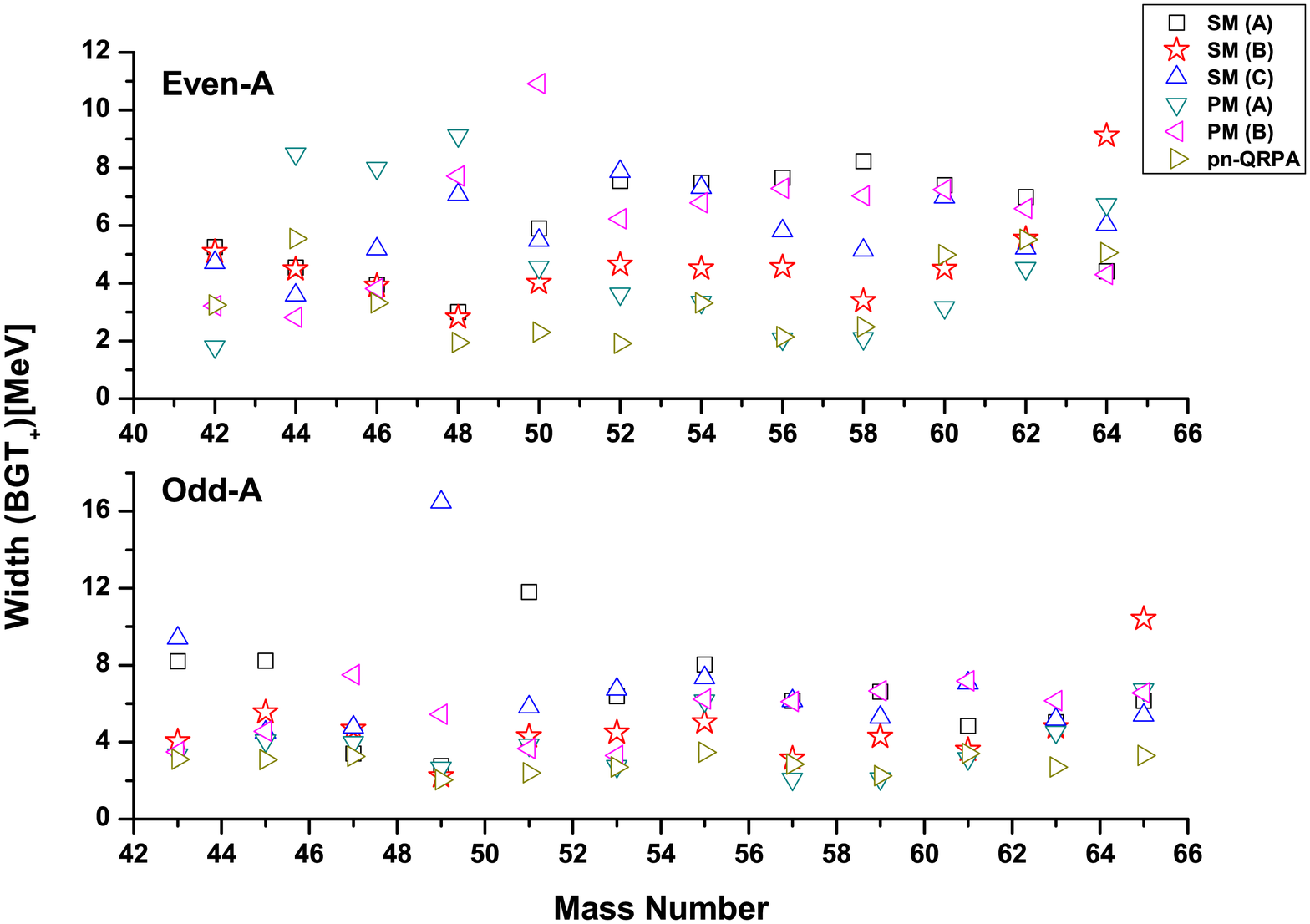}
\end{center}
\caption{Same as Fig.~\ref{fig1} but for width of calculated GT
distributions in electron capture direction.} \label{fig6}
\end{figure}

The widths of calculated GT distribution give an idea how much the
individual GT $1^{+}$ states are scattered about the centroid value.
The widths were calculated for all six models and the results are
shown in Fig.~\ref{fig5} and Fig.~\ref{fig6}. Whereas
Fig.~\ref{fig5} shows the calculated widths, in units of MeV, in the
$\beta$-decay direction, Fig.~\ref{fig6} displays the calculated
widths of all GT distributions in the electron capture direction.
The figures show the fact that the pn-QRPA model calculates low
widths between 2 and 5 MeV in both directions. For calculated GT
distributions in the $\beta$-decay direction, Fig.~\ref{fig5} shows
that the PM~(B) and SM~(C) models calculate bigger widths between 8
and 10 MeV for even-even isotopes. For odd-A cases big widths in
same range are calculated by the SM~(C) and PM~(A) models
(Fig.~\ref{fig5}). The effect of deformation on SM model can be seen
by comparing the SM~(B) and SM~(C) models. For even-even chromium
isotopes it is noted that deformation of nucleus leads to increased
width values except for neutron-deficient and neutron-rich isotopes.
Similarly, all QRPA models in Fig.~\ref{fig6} calculate big widths
except for the pn-QRPA model.

Fig.~\ref{fig1} -- Fig.~\ref{fig6} show the fact that the pn-QRPA
calculated $B(GT)_{-}$ ($B(GT)_{+}$) strength increases (decreases)
monotonically with increasing mass number. Further the pn-QRPA model
calculates low-lying centroid values in daughter and also calculate
smaller widths for the GT distributions. Other models show disparate
results for calculated GT strength distributions. The SM~(C) model
has a tendency to calculate bigger GT strengths, centroids and
widths for the GT distributions. The PM~(B) model generally
calculates low GT strength values.

\begin{figure}
\begin{center}
\includegraphics*[width=15cm,angle=0]{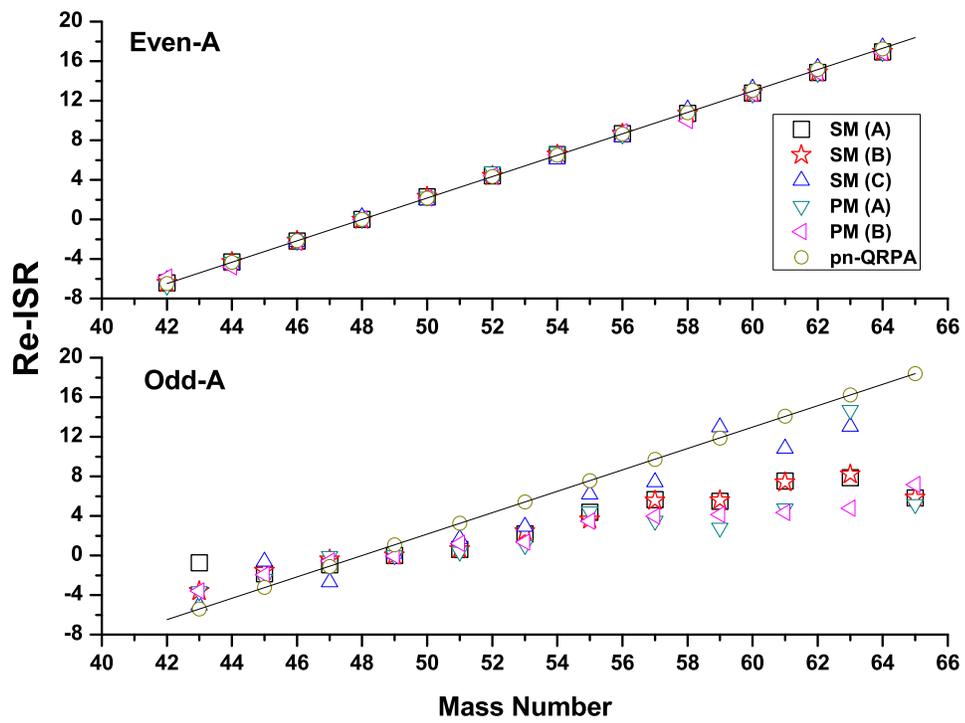}
\end{center}
\caption{Same as Fig.~\ref{fig1} but for comparing the calculated
re-normalized Ikeda sum rule. The straight line is the theoretical
value of the sum rule and is shown just to guide the eye.}
\label{fig7}
\end{figure}

We also wanted to check how our six models perform when it comes to
satisfying the model-independent Ikeda Sum Rule (ISR)
(Eq.~(\ref{Eqt. ISR})). Fig.~\ref{fig7} shows the result. In
Fig.~\ref{fig7} the solid line is the theoretically predicted value
of re-normalized ISR and is shown to guide the eye. Fig.~\ref{fig7}
shows that the re-normalized ISR is obeyed by all models for
even-even cases. For odd-A cases it is only the pn-QRPA model that
obeys Eq.~(\ref{Eqt. ISR}). The SM~(C) model does the best to
satisfy Eq.~(\ref{Eqt. ISR}) among the SM and PM models for odd-A
chromium isotopes. The deviations increase with increasing mass
number for the remaining QRPA models. Perhaps $2p-2h$ configuration
mixing can account for the missing strength in the PM and SM models.
The pn-QRPA model obeyed the ISR.

After the mutual comparison of the six QRPA model calculations of GT
strength distribution for the twenty-four chromium isotopes, we
wanted to see how our calculations compare with other theoretical
models and measurements.  We short-listed three of the best QRPA
models for the sake of this comparison. Amongst the Schematic Model
we chose the SM~(C) model. The choice was rather straight forward as
this model was performing GT calculation in both $pp$ and $ph$
channels and also taking the deformation of nucleus into
consideration. From PM model, PM~(B) was the better choice.
Consequently, the pn-QRPA, SM~(C) and PM~(B) models were
short-listed for the sake of comparison with other work. We searched
the literature for previous calculations/measurements of GT strength
distribution in isotopes of chromium and found a few results. Below
we compare our calculations with these previous findings.

One can find unique information on nuclear structure in the mid-$fp$
shell region by a study of GT decay in $^{46}$Cr. Besides measuring
the favored-allowed GT transition in $^{46}$V, the authors
\citep{On05} also calculated the corresponding GT matrix element
using the shell model. Three different interactions, namely the KB3G
\citep{Pov01}, FPD6 \citep{Ric91} and GXPF2 \citep{Hon02}, were
employed to perform the shell model calculation by the authors. (For
a recent comparison between QRPA and shell model calculated GT
strength distributions, see \citep{Nab13}.) The authors use a
standard quenching factor of $(0.74)^2$ in their calculation. In
Table~\ref{table1}, we compare our calculated $B(GT)_+$ values for
$^{46}$Cr  with the measured and calculated results of Onishi and
collaborators. We also compare our results with the phenomenological
quasideutron (QD) model \citep{Lis01} in Table~\ref{table1}. It can
be seen from Table~\ref{table1} that the closest value for the
measured $B(GT)_+$ to the 1$^{+}_{1}$ state is the value obtained by
shell model using the two body FDP6 interaction. However, our
pn-QRPA calculated value of 0.49 is also very close to the FPD6
value and subsequently to the measured result obtained by Onishi et.
al. The SM~(C) model gives a value of 0.19 for this GT transition.
The calculated total $B(GT)_+$ value using the PM~(B) model is 0.35
up to 9.53 MeV in $^{46}$V. However PM~(B) does not calculate any GT
strength up to 1 MeV and hence was not included in
Table~\ref{table1}. It can be concluded that pn-QRPA model and shell
model using FDP6 interaction give a better description of the
experimental result for total $B(GT)_+$ value of $^{46}$Cr compared
with other calculations.

\begin{figure}
\begin{center}
\includegraphics*[width=10cm,angle=0]{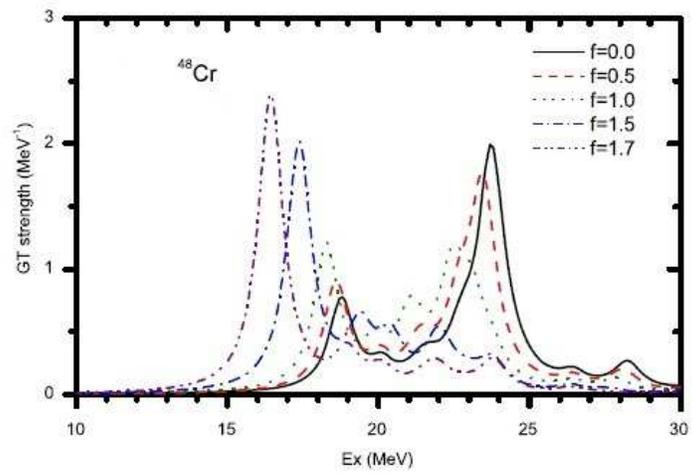}
\end{center}
\caption{ HFB+QRPA calculation reproduced from \citep{Bai13}}
\label{fig8}
\end{figure}

\begin{figure}
\begin{center}
\includegraphics*[width=10cm,angle=0]{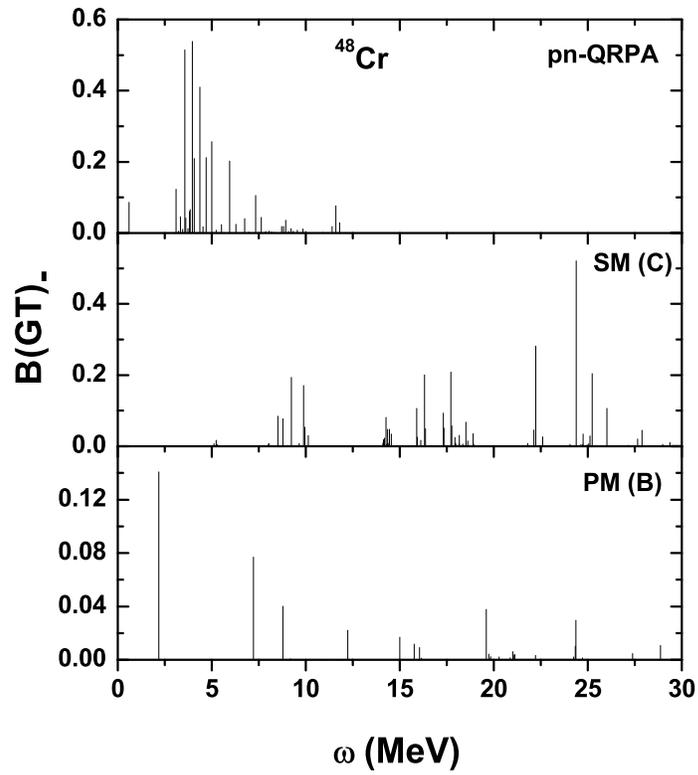}
\end{center}
\caption{Comparison of calculated B(GT)$_{-}$ strength distributions
in $^{48}$Cr among the three chosen QRPA models. $\omega$ represents
excitation energy in $^{48}$Mn in units of MeV.} \label{fig9}
\end{figure}

Bai and collaborators  performed a Hartree-Fock-Bogoliubov + QRPA
calculation using Skyrme interactions for GT states in $N$ = $Z$
nuclei \citep{Bai13}. The idea was to study the role of $T$ = $0$
pairing in the GT states. For this they scaled their pairing
interactions with a factor $f$ which was varied from 0.0 (null
pairing) to 1.7 (strong pairing). Their GT calculation for $^{48}$Cr
is reproduced from their paper and shown as Fig.~\ref{fig8}. The
figure shows that with increase in pairing strength, the centroid of
the GT distribution shifts to lower excitation energy. The authors
calculated a total $B(GT)_{-}$ strength of 11.77. (It is to be noted
that the x-axis of Fig.~\ref{fig8} represents excitation energy in
\textit{parent} nucleus.) The GT strength is distributed roughly
over 15 MeV in parent. On the other hand Fig.~\ref{fig9} shows our
QRPA calculations of GT$_{-}$ strength distribution in $^{48}$Cr. In
all our calculations of GT strength distribution, the x-axis shows
excitation energy in \textit{daughter} nucleus. All calculations
(Fig.~\ref{fig8} and Fig.~\ref{fig9}) show GT fragmentation. Whereas
the pn-QRPA calculates bulk of the GT strength at low excitation
energy in daughter, the SM~(C) model shifts the centroid to higher
excitation energy. In PM~(B) model, the main peak is at around 2.2
MeV , but GT $1^+$ states have been distributed up to 30 MeV in
daughter. High-lying GT states are also seen in the SM~(C) model and
main peaks lie in the energy range of 20-25 MeV. The pn-QRPA model
calculated a total $B(GT)_{-}$ strength of 3.33 and placed the
centroid at 4.78 MeV. The SM~(C) model calculated a similar total
strength of 3.37 but centroid placement was at much higher
excitation energy in $^{48}$Mn (15.4 MeV). On the other hand the
PM~(B) model calculated a very small strength of 0.16 and placed the
energy centroid at 11.5 MeV in daughter.

Exact diagonalization  shell model calculation in the $m$ scheme
using the KB3 interaction  was performed by Caurier and
collaborators \citep{Cau94} for $^{48}$Cr in the electron capture
direction. The authors employed a quenching factor of $(0.77)^2$ in
their calculation. Later Mart\'{i}nez-Pinedo et. al. performed a
similar calculation for $^{49}$Cr \citep{Mar97}. The authors
reported the value of 4.13 for total $B(GT)_+$ of $^{49}$Cr in their
paper. Further the gross properties of even-even and $N$ = $Z$
nuclei with mass number 48 to 64 were studied using the shell model
Monte Carlo (SMMC) methods by Langanke and collaborators
\citep{Lan95}. Calculations were not done for odd-A and odd-odd
nuclei by SMMC because of a sign problem introduced by the
projection of an odd number of particles.  A quenching factor of
$(0.80)^2$ was used in the calculation of total $B(GT)_+$ for
$^{48}$Cr using the SMMC method. Our QRPA results are presented and
compared with these calculations in Table~\ref{table2}. The biggest
value, for both total $B(GT)_+$ and $B(GT)_-$ in $^{48}$Cr, is given
by SMMC. The SMMC and Shell Model results are very close to each
other. Our QRPA calculations give lower B(GT) values which we
attribute to a bigger quenching factor used in our models. For
$^{49}$Cr, the PM~(B) model calculates the lowest strength. Shell
Model did not report the total $B(GT)_-$ strength for $^{49}$Cr.

\begin{figure}
\begin{center}
\includegraphics*[width=10cm,angle=0]{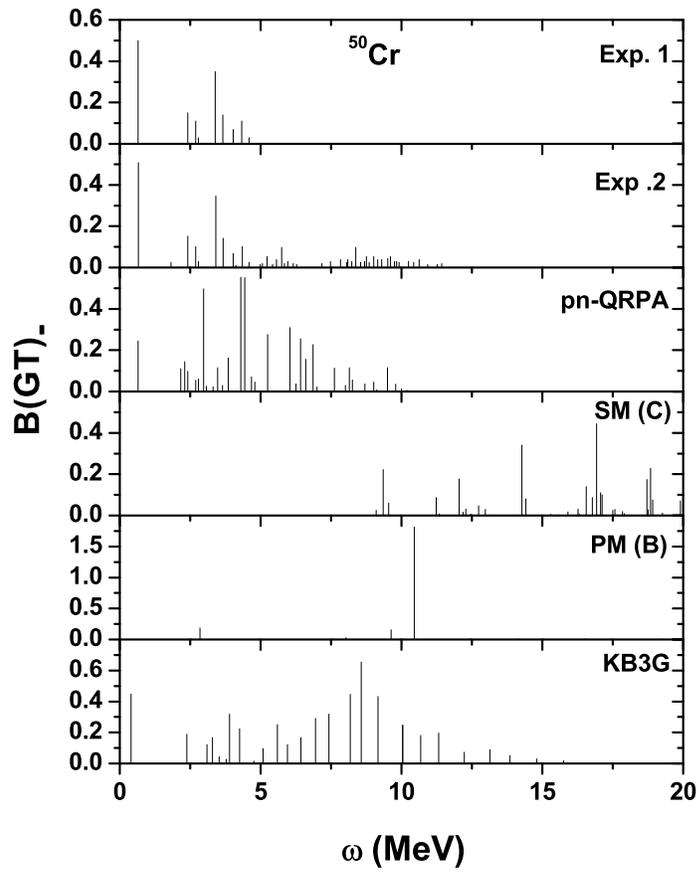}
\end{center}
\caption{Comparison of calculated B(GT)$_{-}$ strength distributions
in $^{50}$Cr with measurements and other theoretical model. Exp. 1
shows measured values by \citep{Fuj11}, Exp. 2 by \citep{Ada07}
while KB3G shows shell model calculation by \citep{Pet07}. $\omega$
represents excitation energy in $^{50}$Mn in units of MeV.}
\label{fig10}
\end{figure}

As mentioned earlier there were two different measurements of
$B(GT)_-$ strength for $^{50}$Cr in literature \citep{Fuj11, Ada07}.
Petermann and collaborators \citep{Pet07} performed large-scale
shell model (LSSM) calculation of the GT strength distributions in
$N \sim Z$ $fp$-shell nuclei (including $^{50}$Cr) using the KB3G
residual interaction \citep{Pov01}. A quenching factor of $(0.74)^2$
was used in the LSSM calculation. In Fig.~\ref{fig10}, we show the
two experimental results, our model calculations and the calculation
by Petermann et. al. for $B(GT)_{-}$ strength distributions in
$^{50}$Cr. Fragmentation of GT $1^+$ states exist in all cases
except in the GT distribution calculated by our PM~(B) model. The
pn-QRPA calculated strength distribution is in very good agreement
with the measured data. In the SM~(C), GT strength appears at higher
energies. There exist only one main peak in the PM~(B) model. GT
$1^+$ states are mainly populated within energy interval of 3-7 MeV
in both pn-QRPA and Exp. 2. The KB3G calculation of Petermann et.
al. shows a wide spectrum of GT distribution covering states up to
15 MeV in daughter. The pn-QRPA placed the centroid of the
calculated $B(GT)_{-}$ strength distribution at 4.81 MeV in
daughter. The corresponding values calculated by the SM~(C) and
PM~(B) models are 19.16 MeV and 2.79 MeV, respectively.
Fig.~\ref{fig10} shows that the pn-QRPA calculation has the best
agreement with the available experimental data.

Table~\ref{table3} presents the comparison of our calculated total
$B(GT)_{-}$ values for $^{50}$Cr with the measured data and LSSM. In
this table the second column gives total $B(GT)_{-}$ value up to 5
MeV in $^{50}$Mn whereas the last column gives the strength up to 12
MeV in daughter. It can be seen from Table~\ref{table3} that LSSM
results are in very good agreement with the measured data up to 5
MeV.  The PM~(B) gives a good description of the experimental data
for excitation energies up to 12 MeV.  In this case the pn-QRPA and
LSSM calculations give bigger total $B(GT)_{-}$ values than the
measured data.

We compare the total $B(GT)_{+}$  for $^{50}$Cr with other
theoretical calculations in Table~\ref{table4}. Large scale
$0\hbar\omega$ shell model calculations for $fp$-shell nuclei were
performed by Caurier et. al. \citep{Cau95} using the KB3 residual
interaction. A quenching factor of $(0.77)^2$ was used in the
calculation. Results for total $B(GT)_{+}$  for $^{50}$Cr along with
other $fp$-shell nuclei were shown in Fig.~2 of \citep{Cau95}. The
SMMC calculated value was taken from \citep{Lan95}. From
Table~\ref{table4} one notes that the Shell Model and SMMC results
are very close to each other. Our calculated total $B(GT)_{+}$
values are lower than the corresponding LSSM and Shell Model results
which we again attribute to the bigger quenching factor used in our
calculation. The SM~(C) and pn-QRPA models calculate similar
strength values for $^{50}$Cr. The PM~(B) gives the lowest value for
total $B(GT)_{+}$ in $^{50}$Cr.

\begin{figure}
\begin{center}
\includegraphics*[width=10cm,angle=0]{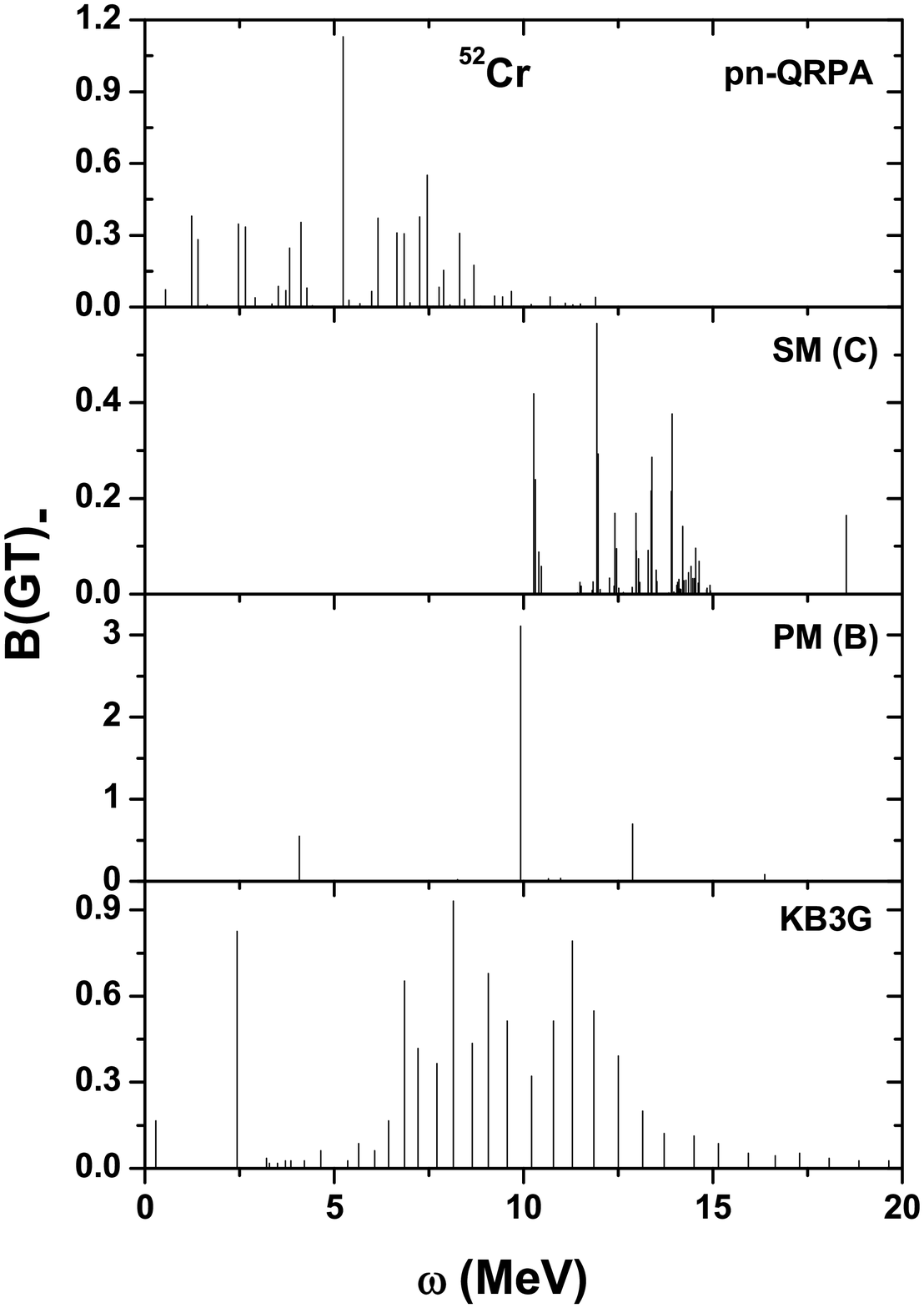}
\end{center}
\caption{Comparison of calculated B(GT)$_{-}$ strength distributions
in $^{52}$Cr with shell model calculation \citep{Pet07}. $\omega$
represents excitation energy in $^{52}$Mn in units of MeV.}
\label{fig11}
\end{figure}

Petermann and collaborators \citep{Pet07} also performed a large
scale shell model calculation of $B(GT)_{-}$ in $^{52}$Cr using the
KB3G interaction. We compare their results with our QRPA
calculations in Fig.~\ref{fig11}. The figure displays the fact that
in this particular case, the pn-QRPA and shell model calculations
produce a better fragmentation of GT $1^+$ states than the ones by
PM~(B) and SM~(C) models. In shell model calculation the GT states
are mainly concentrated between 5--15 MeV in daughter. The PM~(B)
model once again fails to produce the required fragmentation in GT
states and produce a main peak at 9.9 MeV. In the SM~(C) model, GT
$1^+$ stated are calculated  at higher excitation energy (around
10-15 MeV). The pn-QRPA once again places the energy centroid at low
excitation energy of 5.41 MeV in $^{52}$Mn as against the value of
15.54 MeV and 9.88 MeV calculated by SM~(C) and PM~(B) models,
respectively.

Our results for total $B(GT)$ strength (in both directions) for
$^{52,53,54}$Cr are compared with other theoretical models in
Table~\ref{table5}. Nakada and Sebe \citep{Nak96} performed a
large-scale shell model calculation for $fp$-shell nuclei by using
Towner's microscopic parameters and using the KB residual
interaction. The authors incorporated a quenching factor of 0.68 in
their Shell Model calculation. The results are shown in last row of
Table~\ref{table5}. The Shell Model calculates biggest value for
total $B(GT)_{-}$ in $^{52,53,54}$Cr.  The same model also calculate
biggest value for total $B(GT)_{+}$ in $^{52,53}$Cr. SMMC
\citep{Lan95} calculated total $B(GT)_{+}$ only for even-even
nuclei. They were unable to perform calculations for odd-A nuclei
for reasons mentioned earlier. The large scale shell model
calculations of total $B(GT)_{-}$ in $^{52,54}$Cr were also reported
by Petermann and collaborators \citep{Pet07} and are in excellent
agreement with results of our SM~(C) model. For $^{53}$Cr isotope,
we see that the results of the SM~(C) and Shell Model for
$B(GT)_{+}$ are very close to each other. In case of $^{54}$Cr, the
SM~(C) model calculates the biggest value of total $B(GT)_{+}$.
Further, it is noted that the SMMC, pn-QRPA and Shell Model results
are close to each other. It is clear from Table~\ref{table5} that
the Shell Model results of Nakada and Sebe are way too big and
requires further quenching of strength in $\beta$-decay direction.

\begin{figure}
\begin{center}
\includegraphics*[width=10cm,angle=0]{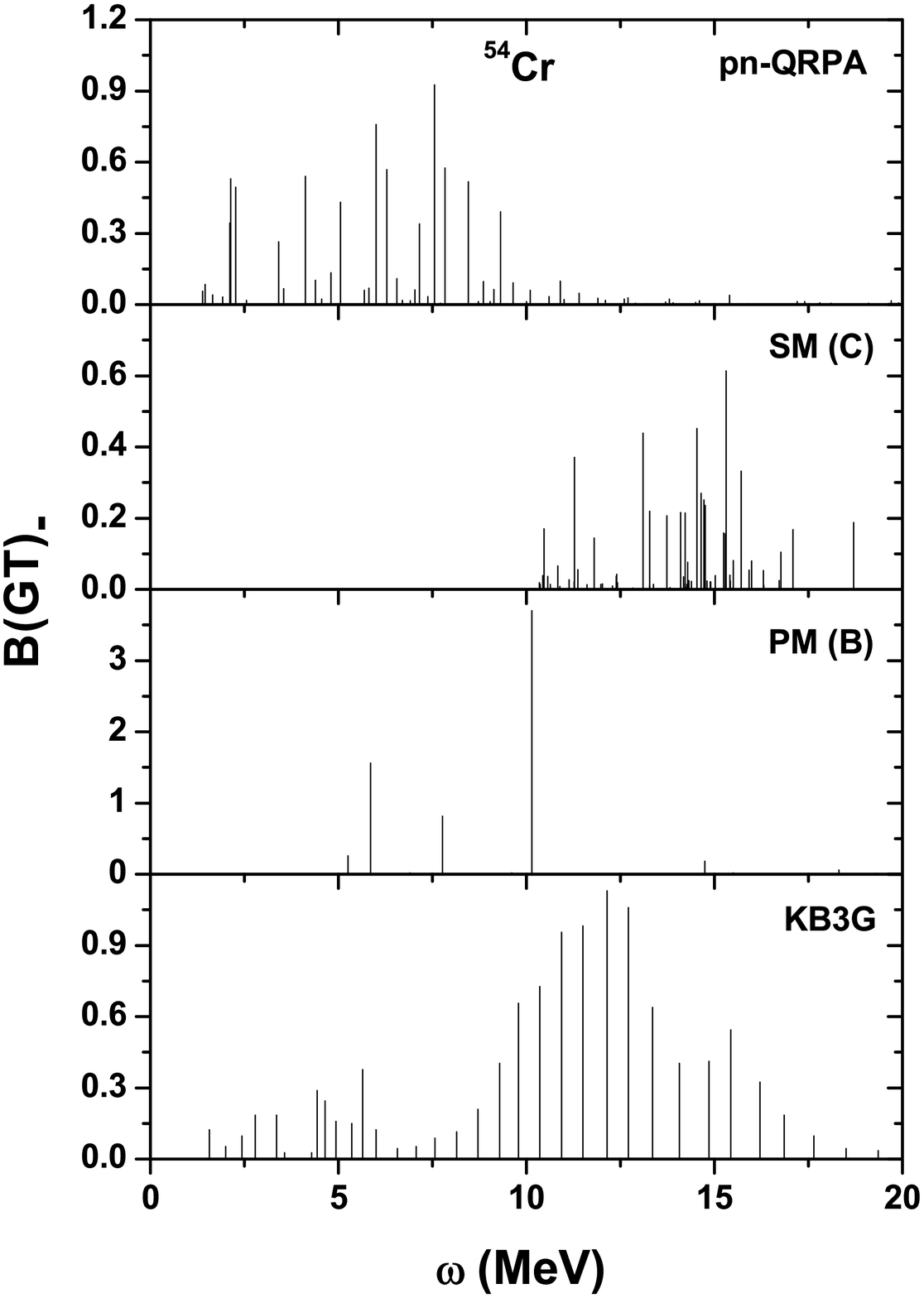}
\end{center}
\caption{Comparison of calculated B(GT)$_{-}$ strength distributions
in $^{54}$Cr with shell model calculation \citep{Pet07}. $\omega$
represents excitation energy in $^{54}$Mn in units of MeV.}
\label{fig12}
\end{figure}

In Fig.~\ref{fig12}, GT strength distributions for $\beta$-decay
transitions in $^{54}$Cr have been presented and compared with the
large scale shell model calculation of Petermann et al. All models
display different degree of GT fragmentation over many daughter
states. Bulk of GT strength in $1^+$ states have been concentrated
in different energy ranges in different models. They are placed at
energy intervals of 2.5--10 MeV in pn-QRPA model , 10--20 MeV in
SM~(C) model and 8--16 MeV in shell model calculation. The PM~(B)
model shows very poor fragmentation. We attribute this to the
neglect of deformation in the model. The assumption of spherical
nuclei in PM~(B) model leads to a concentration of most of the
strength in one particular state. The total $B(GT)_{-}$ strength
calculated by PM~(B) model is 6.68. This is to be compared with the
values of 8.45 and 11.19 calculated by the pn-QRPA and SM~(C)
models, respectively. The pn-QRPA model places the centroid at 6.10
MeV in
 $^{54}$Mn whereas the SM~(C) model places the centroid at more than
 twice this energy.

We finally compare our calculated total $B(GT)_{+}$ values for
$^{56}$Cr with SMMC model and results are shown in
Table~\ref{table6}. For this even-even isotope of chromium, the SMMC
and pn-QRPA results are in very good agreement. The SM~(C) model
calculates the highest value whereas PM~(B) bagged a paltry strength
of 0.09.

\section{Summary and conclusions}
From various studies of the presupernova evolution of massive stars,
it was concluded that weak-interaction mediated reactions in
chromium isotopes (along with other $fp$-shell nuclei) play a key
role in the notorious and complex dynamics of supernova explosion.
For more results concerning weak rates of other key $fp$-shell
nuclei using the pn-QRPA model, see \citep{Nab05, Nab07, Nab08b,
Nab09, Nab10, Nab11, Nab12, Nab13}. Owing to the evidence of strong
connection between supernova explosion and the $r$-process
nucleosynthesis (see e.g. \citep{Cow04}), we felt motivated to
calculate and study the GT transitions in isotopes of chromium.

Six different QRPA models were used to calculate and study the GT
transitions in twenty-four (24) isotopes of chromium. These included
many important neutron-deficient and neutron-rich nuclei. The idea
was to study the theoretical uncertainty involved in the QRPA
models. The different models studied the effect of particle-particle
interaction, particle-hole interaction and deformations in QRPA
calculations. It was concluded that PM~(B) calculated the lowest
B(GT) strength values and failed to produce the desired
fragmentation of the strength. As discussed before, the neglect of
deformation in PM~(B) model led to a concentration of most of the
strength in one particular state. Interaction with a deformed mean
field leads to fragmentation of the GT $1^{+}$ state and of GT
strength \citep{Hir93, On05}. The incorporation of deformation in
the PM~(B) model might improve the situation and we are currently
working on this. The PM~(B) and SM~(C) models did satisfy Ikeda Sum
Rule (ISR) for even-even cases but posed some deviations for the
case of odd-A chromium isotopes. The SM~(C) model calculated biggest
values of total GT strength but at the same time also placed the GT
centroid at high excitation energy in daughter nucleus. In the PM,
the inclusion of $pp$ interaction resulted in higher placement of
centroid values in daughter nuclei. Similarly the inclusion of
deformation in the SM led to calculation of bigger values of
centroid. The pn-QRPA model displayed a tendency to calculate lower
centroids which can translate to bigger weak rates in stellar
environment. The SM and PM models satisfy ISR for even-even isotopes
of chromium. The pn-QRPA model satisfied the ISR for both even-even
and odd-A cases.

Our calculations were also compared with previous measurements and
theoretical calculations wherever available. All our models showed
decent comparison with previous measurements/calcuations in relevant
cases. Of special mention is the very good agreement of pn-QRPA
result with the measured $B(GT)_{-}$ strength distributions in
$^{50}$Cr. It is hoped that the predicted GT strength distributions
for the neutron-rich and neutron-deficient chromium isotopes would
prove very useful for core-collapse simulators and other related
network calculations.

\vspace{0.5 in}\textbf{Acknowledgments:}   J.-U. Nabi wishes to
acknowledge the support provided by T\"{u}bitak (Turkey) under
Project No. 1059B211402772 and the Higher Education Commission
(Pakistan) through the HEC Project No. 20-3099. The authors also
wish to acknowledge the useful discussions  with Cevad Selam on the
results of PM and SM models.

\clearpage


\clearpage
\begin{table}
\caption{Comparison of SM~(C) and pn-QRPA calculated total
$B(GT)_{+}$ values in $^{46}$Cr, up to 1 MeV in daughter nucleus,
with measurement and other theoretical models (Exp, FPD6, KB3, GXPF2
\citep{On05} and QD \citep{Lis01}).}\label{table1}
\begin{tabular}{c|c}
Models &$\sum B(GT)_{+}$\\
\hline
SM~(C)  &0.19\\
pn-QRPA &0.49\\
Exp. &0.64$\pm$0.20 \\
FPD6&0.51 \\
KB3& 0.31\\
GXPF2&0.35 \\
QD& 0.41\\
\end{tabular}
\end{table}

\begin{table}
\caption{Comparison of calculated total $B(GT)$ values in
$^{48,49}$Cr with other theoretical models (SMMC (KB3) \citep{Lan95}
and Shell Model (KB3) \citep{Mar97,Cau94}).}\label{table2}
\begin{tabular}{c|cc|cc}
Models &$^{48}$Cr&&$^{49}$Cr&\\
&$\sum B(GT)_{-}$&$\sum B(GT)_{+}$&$\sum B(GT)_{-}$&$\sum B(GT)_{+}$  \\
\hline
SM~(C) &3.37&3.12&3.33&3.14\\
PM~(B) &0.16&0.16&0.24&0.32\\
pn-QRPA &3.33&3.33&3.31&2.23\\
SMMC (KB3) &4.37$\pm$0.35&4.37$\pm$0.35&-&- \\
Shell Model (KB3) &4.13&4.13&-& 4.13\\
\end{tabular}
\end{table}

\begin{table}
\caption{Comparison of calculated total $B(GT)_{-}$ value in
$^{50}$Cr with measurements and other theoretical models (Exp. 1
\citep{Fuj11}, Exp. 2 \citep{Ada07}, LSSM (KB3G) \citep{Pet07}). The
second column gives total GT strength value up to 5 MeV in daughter
nucleus and the right column up to 12 MeV.}\label{table3}
\begin{tabular}{c|cc}
Models &$^{50}$Cr & \\
&$\sum B(GT)_{-}$ &$\sum B(GT)_{-}$  \\
\hline
SM~(C) &-&0.41\\
PM~(B) &0.18&2.19\\
pn-QRPA &2.79&4.65\\
Exp. 1 &1.49&-\\
Exp. 2 &-&2.69\\
LSSM (KB3G) &1.53&5.20\\
\end{tabular}
\end{table}

\begin{table}
 \caption{Comparison of calculated total $B(GT)_{+}$ value in
$^{50}$Cr with other theoretical models (Shell Model (KB3)
\citep{Cau95} and SMMC (KB3) \citep{Lan95}).}\label{table4}
\begin{tabular}{c|c}
Models &$^{50}$Cr  \\
&$\sum B(GT)_{+}$  \\
\hline
SM~(C)&2.61\\
PM~(B) &0.12\\
pn-QRPA &2.49\\
Shell Model (KB3) &3.57 \\
SMMC (KB3) &3.51$\pm$0.27 \\
\end{tabular}
\end{table}

\begin{table}
\caption{Comparison of calculated total $B(GT)$ values in
$^{52,53,54}$Cr with other theoretical models (LSSM (KB3G)
\citep{Pet07}, SMMC (KB3) \citep{Lan95} and Shell Model (KB)
\citep{Nak96}).}\label{table5} \scriptsize
\begin{tabular}{c|cc|cc|cc}
Models&$^{52}$Cr&&$^{53}$Cr&&$^{54}$Cr\\
&$\sum B(GT)_{-}$&$\sum B(GT)_{+}$&$\sum B(GT)_{-}$&$\sum B(GT)_{+}$ &$\sum B(GT)_{-}$&$\sum B(GT)_{+}$ \\
\hline
SM~(C) &8.14&3.78&6.57&3.66&11.19&5.12\\
PM~(B) &4.57&0.18&1.46&0.12&6.68&0.17\\
pn-QRPA &6.55&2.21&5.91&0.51&8.45&1.95\\
LSSM (KB3G) &8.85&-&-&-&11.13&- \\
SMMC (KB3) &-&3.51$\pm$0.19&-&-&-&2.21$\pm$0.22 \\
Shell Model (KB) &17.4&4.3&20.1&3.8&22.4&2.9 \\
\hline
\end{tabular}
\end{table}

\begin{table}
\caption{Comparison of calculated total $B(GT)_{+}$ values in
$^{56}$Cr with SMMC (KB3) \citep{Lan95}.}\label{table6}
\begin{tabular}{c|c}
Models &$^{56}$Cr\\
&$\sum B(GT)_{+}$  \\
\hline
SM~(C) &5.32\\
PM~(B) &0.09\\
pn-QRPA &1.31\\
SMMC (KB3)&1.5$\pm$0.21 \\
\hline
\end{tabular}
\end{table}


\begin{thebibliography}{}

\bibitem[Adachi et al.(2007)]{Ada07} Adachi T.,  et al.
Gamow-Teller transitions in pf -shell nuclei studied in ($^3$He, t)
reactions. Nucl. Phys. A 788, 70c-75c, 2007.
\bibitem [Audi et al.(2003)]{Aud03}Audi, G., Wapstra, A.H., Thibault, C. The AME2003
Atomic Mass Evaluation (II). Tables, Graphs and References. Nucl.
Phys. A 729, 337-676, 2003.
\bibitem[Aufderheide et. al.(1994)]{Auf94} Aufderheide, M.B., Fushiki, I.,
Woosley, S.E., Stanford, E. and Hartmann, D.H.  Search for important
weak interaction nuclei in presupernova evolution. Astrophys. J.
Suppl. Ser. 91, 389-417, 1994.
\bibitem[Aufderheide et. al.(1996)]{Auf96} Aufderheide, M.B., Bloom, S.D., Mathews, G.J. and
Resler, D.A. Importance of (n,p) reactions for stellar beta decay
rates. Phys. Rev. C 53, 3139-3142, 1996.

\bibitem [Babacan et al.(2004)]{Bab04} Babacan, T., Salamov, D.I. and Kucukbursa, A. The effect of the pairing
interaction on the energies of isobar resonanse in $^{112-124}$Sb
and isospin admixture in $^{100-124}$Sn isotopes. J. Phys. G 30,
759-770, 2004.
\bibitem [Babacan et al.(2005)]{Bab05} Babacan, T., Salamov, D.I. and Kucukbursa, A. Gamow-Teller $1^{+}$ states
in $^{208}$Bi. Phys. Rev. C 71, 037303, 2005.
\bibitem [Babacan et al.(2005a)]{Bab05a}Babacan T., Salamov, D. I. and Kucukbursa, A. The investigation of the
log(ft) values for the allowed Gamow-Teller transitions of some
deformed nuclei. Math. Comp. Appl. 10, 359-368, 2005.
\bibitem[Bai et al.(2013)]{Bai13} Bai, C.L., Sagawa, H., Sasano, M., Uesaka, T., Hagino, K., Zhang,
H.Q., Zhang,  X.Z. and Xu, F.R. Role of T = 0 pairing in
Gamow–Teller states in N = Z nuclei. Phy. Lett. B 719, 116-120,
2013.


\bibitem [Cakmak et al.(2010)]{Nec10} Cakmak, N., Unlu, S. and Selam, C. Gamow-Teller $1^{+}$ states in $^{112-124}$Sb
isotopes, Pramana J. Phys., 75, 649-663, 2010.
\bibitem[Cakmak et. al.(2014)]{Cak14}Cakmak, S., Nabi, J.-U.,
Babacan, T. and Selam, C. Study of Gamow–Teller transitions in
isotopes of titanium within the quasi particle random phase
approximation. Astrophys. Space Sci. 352, 645-663, 2014.
\bibitem[Caurier et al.(1994)]{Cau94} Caurier, E., Zuker, A.P.,
Poves, A., and Mart\'{i}nez-Pinedo, G. Full pf shell model study of
A=48 nuclei. Phys. Rev. C 50, 225-236, 1994.
\bibitem[Caurier et al.(1995)]{Cau95} Caurier, E., Mart\'{i}nez-Pinedo, G., Poves,
A., Zuker, A.P. Gamow-Teller strength in $^{56}$Fe and $^{56}$Fe
Phys. Rev. C 52, R1736-R1740, 1995.
\bibitem[Cowan $\&$ Thielemann(2004)]{Cow04} Cowan, J.J., Thielemann,
F.-K. R-process Nucleosynthesis in Supernovae. Phys. Today Oct. 2004
issue, 57/10 47, 2004.




\bibitem[Fujita et al.(2011)]{Fuj11} Fujita, Y., Rubio, B.,
Gelletly, W. Spin-isospin excitations probed by strong, weak and
electro-magnetic interactions. Prog. Part. Nucl. Phys. 66, 549-606,
2011.
\bibitem[Fuller et al.(1980,1982,1985)]{Ful85} Fuller, G.M., Fowler, W.A. and Newman, M.J. Stellar Weak-Interaction Rates for sd-Shell
Nuclei. I. Nuclear Matrix Element Systematics with Application to
$^{26}$Al and Selected Nuclei of Important to the Supernova Problem.
Astrophys. J. Suppl. Ser. 42, 447-473, 1980; Stellar
Weak-Interaction Rates for sd-Shell Nuclei. I. Nuclear Matrix
Element Systematics with Application to 26Al and Selected Nuclei of
Important to the Supernova Problem. Astrophys. J. Suppl. Ser. 48,
279-320, 1982; Neutron Shell Blocking of Electron Capture During
Gravitational Collapse. Astrophys. J. 252, 741-764, 1982; Stellar
Weak Interaction Rates for Intermediate Mass Nuclei. IV.
Interpolation Procedures for Rapidly Varying Lepton Capture Rates
Using Effective log (ft)- Values. Astrophys. J. 293, 1-16, 1985.

\bibitem[Heger et. al.(2001)]{Heg01} Heger, A., Woosley, S.E., Mart\'{i}nez-Pinedo,  G. and
 Langanke, K.  Presupernova evolution with improved rates for weak
interactions. Astrophys. J. 560, 307-325, 2001.
\bibitem [Hirsch et al.(1993)]{Hir93} Hirsch, M., Staudt, A., Muto, K. and Klapdor-Kleingrothaus, H.V.
Microscopic predictions of $\beta^{+}$/EC-decay half-lives. At. Data
Nucl. Data Tables 53, 165-193, 1993.
\bibitem[Honma et al.(2002)]{Hon02}
 Honma, M., Otsuka, T., Brown, B.A. and  Mizusaki, T. Effective
interaction for pf-shell nuclei. Phys. Rev. C 65, 061301(R), 2002.


\bibitem[Ikeda et al.(1963)]{Ike63} Ikeda, K.I., Fuji, S.,
Fujita, J.I. The (p,n) reactions and beta decays. Phys. Lett. 3,
271-272, 1963.


\bibitem[Langanke et al.(1995)]{Lan95} Langanke, K., Dean, D.J.,
Radha, P.B., Alhassaid,  Y.and Koonin, S.E. Shell-model Monte Carlo
studies of fp-shell nuclei. Phys. Rev. C 52, 718-725, 1995.
\bibitem[Lisetskiy et al.(2001)]{Lis01} Lisetskiy, A.F., Jolos, R.V., Pietralla, N. and
Brentano, P. von. Quasideuteron configurations in odd-odd N=Z
nuclei. Phys. Rev. C 60, 064310, 1999; Lisetskiy, A.F., Gelberg, A.,
Jolos, R.V., Pietralla, N. and Brentano, P. von. Quasideuteron
states with deformed core. Phys. Lett. B 512, 290-296, 2001.


\bibitem[Mart\'{i}nez-Pinedo et al.(1997)]{Mar97} Mart\'{i}nez-Pinedo, G., Zuker, A.P., Poves , A. and
Caurier, E. Full pf shell study of A=47 and A=49 nuclei. Phys. Rev.
C 55, 187-205, 1997.
\bibitem [M\"{o}ller et al.(1981)]{Moe81}M\"{o}ller, P. and Nix, J.R. Atomic Masses and Nuclear
Ground-State Deformations Calculated with a New
Macroscopic-Microscopic Model. At. Data Nucl. Data Tables 26,
165-196, 1981.

\bibitem [Nabi $\&$ Klapdor-Kleingrothaus(1999)]{Nab99}  Nabi, J.-Un and Klapdor-Kleingrothaus, H.V. Weak
Interaction Rates of sd-Shell Nuclei in Stellar Environments
Calculated in the Proton-Neutron Quasiparticle Random-Phase
Approximation. At. Data Nucl. Data Tables  71, 149-335, 1999.
\bibitem [Nabi $\&$ Klapdor-Kleingrothaus(1999a)]{Nab99a}Nabi, J.-Un and Klapdor-Kleingrothaus, H.V.  Microscopic
calculations of weak interaction rates of nuclei in stellar
environment for A = 18 to 100. Eur. Phys. J. A 5, 337-339, 1999.
\bibitem[Nabi $\&$ Klapdor-Kleingrothaus(2004)]{Nab04}Nabi, J.-Un and Klapdor-Kleingrothaus, H.V.  Microscopic
calculations of stellar weak interaction rates and energy losses for
fp- and fpg-shell nuclei. At. Data Nucl. Data Tables 88, 237-476,
2004.
\bibitem[Nabi $\&$ Rahman(2005)]{Nab05} Nabi, J.-Un and Rahman, M.-Ur.  Gamow-Teller strength
distributions and electron capture rates for $^{55}$Co and
$^{56}$Ni. Phys. Lett. B612, 190-196, 2005.
\bibitem [Nabi et al.(2007)]{Nab07}  Nabi, J.-Un, Sajjad , M. and Rahman, M.-Ur.
Electron capture rates on titanium isotopes in stellar matter. Acta
Physica Polonica B 38, 3203-3223, 2007.
\bibitem [Nabi $\&$ Sajjad(2008)]{Nab08b} Nabi, J.-Un and Sajjad, M. Neutrino energy loss rates
and positron capture rates on $^{55}$Co for presupernova and
supernova physics. Phys. Rev. C 77, 055802, 2008.
\bibitem [Nabi(2009)]{Nab09} Nabi, J.-Un. Weak-interaction-mediated rates on
iron isotopes for presupernova evolution of massive stars. Eur.
Phys. J. A 40, 223-230, 2009.
\bibitem [Nabi(2010)]{Nab10}  Nabi, J.-Un. Expanded calculation of neutrino cooling rates due to $^{56}$Ni in stellar matter.
Phys. Scr. 81, 025901, 2010.
\bibitem [Nabi(2011)]{Nab11} Nabi, J.-Un. Ground and excited states Gamow-Teller strength distributions of iron isotopes and
associated capture rates for core-collapse simulations. Astrophys
Space Sci. 331, 537-554 (2011).
\bibitem [Nabi(2012)]{Nab12}Nabi, J.-Un. Nickel isotopes in stellar matter. Eur.
 Phys. J. A, 48, 84, 2012.
\bibitem [Nabi $\&$ Johnson(2013)]{Nab13}Nabi, J.-Un and Johnson, C.W. Comparison of Gamow-Teller strengths in the random phase
approximation. J. Phys. G 40, 065202, 2013.
\bibitem[Nakada $\&$ Sebe(1996)]{Nak96} Nakada, H and Sebe T. Microscopic description of Gamow–Teller transitions in
middle pf-shell nuclei by a realistic shell-model calculation. J.
Phys. G: Nucl. Part. Phys. 22, 1349–-1362, 1996.

\bibitem[Onishi et al.(2005)]{On05}
 Onishi, T.K., Gelberg, A., Sakurai, H., Yoneda, K., Aoi,  N., Imai, N.,
Baba, H., Brentano, P. von, Fukuda, N. , Ichikawa, Y., Ishihara, M.,
Iwasaki, H., Kameda, D., Kishida,T.,  Lisetskiy, A.F., Ong, H.J.,
Osada, M., Otsuka, T., Suzuki, M.K., Ue, K., Utsuno, Y.  and
Watanabe, H. Gamow-Teller decay of the T=1 nucleus $^{46}$Cr. Phys.
Rev. C 72, 024308, 2005.

\bibitem[Petermann et al.(2007)]{Pet07}
 Petermann, I., Mart\'{i}nez-Pinedo, G.,  Langanke, K. and Caurier, E.
Breaking of the SU(4) limit for the Gamow-Teller strength in N
$\sim$ Z nuclei. Eur. Phys. J. A 34, 319–-324, 2007.
\bibitem[Pyatov $\&$ Salamov(1977)]{Pya77} Pyatov, N.I. and  Salamov, D.I. Conservation laws and collective
excitations in nuclei. Nucleonica 22, 1-127, 1977.
\bibitem[Poves et al.(2001)]{Pov01}
 Poves, A., Sánchez-Solano, J., Caurier, E. and Nowacki, F. Shell model
study of the isobaric chains A=50, A=51 and A=52.  Nucl. Phys. A694,
157-198, 2001.

\bibitem[Raman et. al.(1987)]{Ram87} Raman,  S., Malarkey, C.H., Milner, W.T., Nestor, C.W., Stelson, Jr. and P.H.
Transition Probability, B(E2)$\uparrow$, from the Ground to the
First-Excited 2$^{+}$ State of Even-Even Nuclides, At. Data Nucl.
Data Tables, 36, 1--96, 1987.
\bibitem[Richter et al.(1991)]{Ric91}
 Richter, W.A., Merwe, M.G. Van Der,  Julies, R.E. and  Brown,  B.A. New
effective interactions for the $0f1_p$ shell Nucl. Phys. A 253,
325-353, 1991.

\bibitem [Salamov et al.(2003)]{Sal03} Salamov, D.I., Kucukbursa, A., Maras, I., Aygor, H.A.,
Babacan, T. and Bircan, H. Calculation of the Log(ft) Values for the
Allowed Gamow-Teller Transitions in Deformed Nuclei Using the Basis
of Woods-Saxon Wave Functions, Acta Physica Slovaca, 53, 307-319,
2003.
\bibitem [Selam et al.(2004)]{sel04} Selam, C., Babacan, T., Bircan, H., Aygor, H.A., Kucukbursa, A. and , Maras, I.
The investigation of the Log(ft) Values for the Allowed Gamow-Teller
Transitions of Some Deformed Nuclei, Mathematical Computational
Applications, 9 (1), 79-90, 2004.
\bibitem [Salamov et al.(2006)]{Sal06} Salamov, D.I., et. al. Proceedings of 5th Conference on Nuclear and Particle Physics (NUPPAC 05), 361
(Cairo, August 2006)
\bibitem[Staudt et. al.(1990)]{Sta90} Staudt, A., Bender, E., Muto
and Klapdor-Kleingrothaus, H.V.  Second-Generation Microscopic
Predictions of Beta-Decay Half-lives of Neutron-Rich Nuclei. {\it
At. Data Nucl. Data Tables}, 44, 79-132, 1990.
\bibitem [Stetcu $\&$ Johnson(2004)]{Ste04}Stetcu, I. and Johnson, C. W.  Gamow-Teller transitions
and deformation in the proton-neutron random phase approximation.
Phy. Rev C 69, 024311, 2004.


\bibitem[Vetterli et al.(1989)]{Vet89}  Vetterli, M.C., H\"{a}usser, O., Abegg, R., Alford, W.P.,
Celler, A., Frekers, D., Helmer, R., Henderson, R., Hicks, K.H.,
Jackson, K.P., Jeppesen, R.G., Miller, C.A., Raywood, K. and Yen, S.
Gamow-Teller strength deduced from charge exchange reactions on
$^{54}$Fe at 300 MeV. Phy. Rev. C 40, 559-569, 1989.

\bibitem[Zioni et al.(1972)]{Zio72}  Zioni, J., Jaffe, A.A., FriedMan, E.,  Haik, N., Schectman, R. and Nir, D.
An investigation of proton-rich nuclei and other products from the
bombardment of $^{24}$Mg, $^{28}$Si and $^{32}$S by $^{16}$O ions.
Nucl. Phys. A 181, 465-476, 1972.


\end{thebibliography}
\end{document}